\documentclass[aps,pra,10pt,twocolumn,showpacs,superscriptaddress]{revtex4-1}

\usepackage{amsmath}
\usepackage{latexsym}
\usepackage{amssymb}
\usepackage{bm}
\usepackage{graphics,epstopdf}
\usepackage{color}
\usepackage{hyperref}

\usepackage{newlfont}
\usepackage{amsfonts}
\usepackage{amsthm}
\usepackage{graphicx}
\usepackage{epsfig}

\newcommand{\ket}[1]{|{#1}\rangle}

\usepackage{times}

\newcommand{\be}{\begin{equation}}
\newcommand{\ee}{\end{equation}}
\newcommand{\bc}{\begin{center}}
\newcommand{\ec}{\end{center}}
\newcommand{\bea}{\begin{eqnarray}}
\newcommand{\eea}{\end{eqnarray}}
\newcommand{\ba}{\begin{array}}
\newcommand{\ea}{\end{array}}

\usepackage{rotating}
\usepackage{tikz}
													
\usepackage{amsmath}
\usepackage{latexsym}
\usepackage{amssymb}
\usepackage{bm}
\usepackage{graphics,epstopdf}
\usepackage{color}
\usepackage{hyperref}

\usepackage{newlfont}
\usepackage{amsfonts}
\usepackage{amsthm}
\usepackage{graphicx}
\usepackage{epsfig}
\usepackage{times}

\usepackage{caption}
\usepackage{subcaption}

\begin{document}

\title{Reverse Quantum Speed Limit: How Slow  Quantum Battery Can Discharge?}

\author{Brij Mohan}

\author{Arun K. Pati}

\affiliation{Quantum Information and Computation Group,\\
Harish-Chandra Research Institute, HBNI, Chhatnag Road, Jhunsi, Prayagraj 211 019, India\\
}

\begin{abstract}

  We introduce the notion of reverse quantum speed limit for arbitrary quantum evolution which answers a
fundamental question: ``how slow a quantum system can evolve in time?" Using the geometrical approach to quantum mechanics, the  reverse speed limit follows from the fact that the gauge invariant length of the reference section is always greater than the Fubini-Study distance on the projective Hilbert space of the quantum system. We illustrate the reverse speed limit for two-level quantum systems with an external driving Hamiltonian and show that our results hold well. We find several examples where our bound is tight. We also find one  practical application of the reverse speed limit in discharging process of quantum batteries which answers the question: ``how slow quantum batteries can discharge?" Our result provides a lower bound on the average discharging power of quantum batteries.

\end{abstract}

%..........................................................................................................................

\maketitle
\section{Introduction}

	Quantum physics has several inherent limitations. These limitations could be either operational or dynamical limitations. From a dynamical point of view the 
	quantum speed limit (QSL) has played a pivotal role in quantum computation, quantum control, and even in quantum thermodynamics. These limitations are not only 
	crucial for theoretical questions but also have practical relevance, as recent years have witnessed the rapidly developing quantum technologies.

The question of how fast a quantum system can evolve in time was first addressed in Ref. \cite{MandelstamTamm01}.
 The notion of speed of transportation of a quantum system, what is now known as the  `quantum speed limit', was 
  first introduced by Aharonov and Anandan \cite{Anandan02} using the Fubini-Study metric. The speed of transportation of quantum system for unitary and non-unitary evolutions was introduced in Ref. \cite{AKPati03, AKPati04, AKPati05} using the Riemannian metric. The QSL is defined as the maximal dynamical evolution speed of the quantum system. It determines the minimal dynamical evolution time required for a quantum state of a quantum system to evolve from an initial state to target state \cite{MandelstamTamm01,Anandan02,S.Deffner15,Poggi16,Margolus17,Levitin18,Vaidman19,Eberly20,Bauer21,Bhattacharyya22,Leubner23,Gislason24,Uhlmann25,Uffink26,Pfeifer27,Horesh28,AKPati29,Soderholm30,Giovannetti31,Andrecut32,Gray33,Luo34,Batle35,Borras36,Zielinski37,Zander38,Andrews39,Kupferman40,urtsever41,Fu42,
Chau43,Jones44,Zwierz45,Taddei46,Campo47,Fung48,Deffner49,Deffner50,Andersson51,Fung52,D.Mondal53,D.Mondal54}.
  Since QSL determines how fast quantum systems evolve it is natural to ask if there is a reverse speed limit for quantum evolution. To answer this question, we formally introduce the notation of reverse quantum speed limit (RQSL) for arbitrary quantum evolution. This is defined as the minimal evolution speed of closed quantum systems and it sets a bound on the maximal evolution time required for a quantum state of a closed quantum system to evolve from an initial state to a target state. This is an upper bound on quantum evolution time. The quantum evolution will be slower, if reverse quantum speed limit time increases. It may have several meaningful applications in the field of quantum physics ranging from quantum information, quantum computation to quantum optimal control and quantum battery.

Quantum battery (QB) was originally introduced by Alicki and Fannes \cite{R.Alicki06}. It consists of small quantum systems with many degrees of freedoms in which we can store energy or extract energy from them. Quantum batteries are much better than classical batteries in many ways because they offer several quantum mechanical advantages \cite{Andolina02,Andolina01}. In recent years, lots of theoretical models of quantum batteries have been studied by several groups to enhance the feature of quantum batteries by exploiting the non-classical resources of quantum mechanics. Also, several charging and discharging protocols have been proposed to enhance the power \cite{D.Ferraro10,F.C.Binder11,Andolina61,Le13,S.Ghosh,Luis}, work storage \cite{R.Alicki06,Crescente}, stability \cite{Quach,S.Gherardini56,A.C.Santos57}, etc. With the help of QSL, it has been found that how fast we can charge the quantum batteries \cite{X.Zhang07,F.C.Binder11,F.Campaioli12,F.Campaioli58}. Here with the help of RQSL, we will  answer 
the question ``how slow quantum batteries can discharge?" QSL and RQSL  also set the bound on average charging/discharging power of quantum batteries. We believe that it will help us in the practical realization of quantum batteries which can maintain its power for longer duration. Ideally, one should design quantum batteries which discharge slowly while operating and hence our RQSL play a pivotal role in deciding the figure of merit of such quantum batteries.

 Our paper is organised as follows. In section II, we provide a basic framework to appreciate various geometric structures that will be used for proving the reverse 
 quantum speed limit. In section III, we apply RQSL for two-level systems. In section IV, we studied the slow discharging of quantum batteries by employing RQSL.
 Finally, we summarize our results in section V.

 %..........................................................................................................................

\section{GEOMETRICAL REVERSE QUANTUM SPEED LIMIT FOR UNITARY EVOLUTIONS}
Before we prove the reverse quantum speed limit, we need some background on geometry of quantum evolution.

\subsection{For pure initial state}
Let us consider a set of  vectors $\{\psi\}$ of $(D+1)$-dimensional quantum system  that belongs to a Hilbert space ${\cal H}^{D+1}$. If these vectors are not normalized, 
we can consider $ \{ \psi/|| \psi ||  \}$ be a set of vectors of norm one belongs to unit normed Hilbert space $\mathcal{L}$.
The state of a quantum system is represented by a ray in the ray
space ${\cal R} = {\cal L}/U(1)$. Two normalized state vectors $|\psi\rangle$ and $|\psi'\rangle$ are equivalent if they belongs to same ray, i.e., 
they merely differ by a phase factor (${|\psi'\rangle} \equiv {e^{i\phi}|\psi\rangle}$, where $e^{i\phi}$ $\in$   $U(1)$). The set of rays of $\mathcal{H}$ via a projection map 
is known as the projective Hilbert space $\mathcal{P}$. The projection map $\Pi:\mathcal{L} \xrightarrow{} \mathcal{P}$ is a principal fibre bundle  
$\mathcal{L}(\mathcal{P},U(1),\Pi)$, with structure group $U(1)$. This can be observed by considering the action of the multiplicative group $C^{*}$ of non-zero complex numbers 
on the space $C^{D+1}-\{0\}$ given by the
equivalence relation $(z_{1},z_{2},. ..,z_{D+1})\lambda:= (z_{1}\lambda,z_{2}\lambda,. ..,z_{D+1}\lambda)$ $\forall$  $\lambda$  $\in$ $C^{*}$. 
This is a free action and the orbit space is the space $\mathcal{CP}^{D}$  of complex lines in the Hilbert space $\mathcal{H}=\mathcal{C}^{D+1}$. 
Thus, we get 
the principal bundle $C^{*}\xrightarrow{}C^{D+1}-\{0\}\xrightarrow{}\mathcal{CP}^{D}=\mathcal{P}$ in which the projection map associates with each $(D+1)$ tuple $(z_{1},z_{2},. ..,z_{D+1})$ the point in $\mathcal{CP}^{D}$ with the homogeneous coordinates. Thus, any pure quantum state at given instant of time is represented by a point in $\mathcal{P}$ and the evolution of the quantum system is represented by a curve $\Gamma$ in $\mathcal{H}$, which projects to a curve $\hat{\Gamma}$ $=\Pi({\Gamma})$ in $\mathcal{P}$ \cite{AKPati03,AKPati04,AKPati05}.\\

%................................................................
\begin{tikzpicture}

\draw[black,thick] (-3.4,-1) -- (2.5,-1,0) -- (3.4,1,0) -- (-2.5,1,0) -- cycle;

 \draw[black] (-2,1.5) -- (-2,4.4);
\draw[dotted,thick] (-2,0) -- (-2,1.5);

\draw [blue,dotted,thick] (2,0) arc[x radius=2, y radius=.5, start angle=0, end angle=180];

\draw[blue,thick] (-2,0) arc (-180:60:2 and 0.5);

\draw[rotate around={10:(-2,2)}][violet,thick](-2,2) arc[x radius=2, y radius=.5, start angle=-180, end angle=0];

\draw[rotate around={10:(-2,2)},violet,dotted,thick](-2,2) arc[x radius=2, y radius=.7, start angle=180, end angle=0];

x radius=2, y radius=.5, start angle=-180, end angle=50];

\draw[rotate around={10:(-2,2)}][violet,thick]  (2,2) arc[x radius=2, y radius=.7, start angle=0, end angle=55];

\draw[rotate around={-0.2:(-2,2.69)},violet,dotted,thick]  (-2,2.69) arc (0:121:-2 and 0.5);

\node[black] at (-0.4,-1.25) {Projective Hilbert space  $\mathcal{P}$};

\node[black] at (-3.4,3.55) {\small{Ray space $\mathcal{R}$}};
%\node[black] at (-3.3,3.25) {\small{  (ray space)}};

\node[thick,black] at (2.2,0) { $\hat{\Gamma}$};

%\node[black] at (2.4,3.6) {$|\chi(t)\rangle$};

\node[black] at (2.15,2.7) {$\Gamma$};
%\node[black] at (2.4,1.6) { $| {\bar {\psi}}(t) \rangle$};

%\node[black] at (-2.5,0) { $t=0$};
%\node[black] at (1.3,0.7) { $t=T$};
\node[black] at (0,-0.75) {\small{ $\rho(t)=|\psi(t)\rangle \langle \psi(t)|$}};
%\node[black] at (-0.4,0.75) { $\rho(t)_{Geodesic}$};

\filldraw[blue] (1.0,0.425) circle (1pt) ;

\filldraw[blue] (-2,0) circle (1pt) ;
\filldraw[violet,thick] (-2,2.69) circle (1pt) ;
\filldraw[black,thick] (-2,1.98) circle (1pt) ;
\filldraw[violet,thick] (1.0,3.12) circle (1pt) ;

\node[thick,black] at (-2.5,2) {\small{ ${|\psi(0)\rangle}$}};
\node[thick,black] at (-3.4,2.75) {\small{ ${|\psi(T)\rangle}={e^{i \bold{\phi}}|\psi(0)\rangle}$}};

\end{tikzpicture}
%................................................................
\begin{small}
Fig 1: In this schematic diagram the violet line represents the curve $\Gamma$ in $\mathcal{H}$. The curve $\Gamma$ is actual  trajectory of non-cyclic evolution of quantum system whose state vector $|\psi(t)\rangle \in\mathcal{H}$ evolves according to the Schr\"odinger equation. For cyclic evolution the curve $\Gamma$ begins and ends on the same ray but at different points. The blue line represents the curve $\hat \Gamma$ in $\mathcal{P}$. The curve $\hat \Gamma$ is basically projection of curves $\Gamma$ in $\mathcal{P}$. Note here that these may be infinite number of such curves $\Gamma$ in $\mathcal{H}$ which can give rise to the same curve $\hat \Gamma$ in $\mathcal{P}$.
\end{small} 
\\

During a non-cyclic evolution of a quantum system, the initial state and the final state belong to two different rays of the Hilbert space. Thus, the evolution curve $\hat{\Gamma}$ is an open path in $\mathcal{P}$ where the initial and final points lie on two different rays.
Using the Pancharatnam connection \cite{Pancharatnam60}, we can compare the relative phases of state vectors belonging to two different rays. If a quantum system evolves from an  initial state to the final state, then the relative phase difference between these states is given by
\begin{equation}
e^{i\Phi} = \frac{\langle\psi(0 )|\psi(t)\rangle}{|\langle\psi(0 )|\psi(t)\rangle|}.
\end{equation}
Here,  the initial and final states should not be orthogonal.
If $\langle\psi(t)|\psi(0)\rangle$ is complex, then the quantum system does acquire relative phase during the evolution of the system. In this case if we map the open path $\hat{\Gamma}$ in  $\mathcal{P}$ to $\mathcal{L}$, there are  many open curve in $\mathcal{L}$ corresponding to this open curve in $\mathcal{P}$. Among all of them, there exists one special open curve, which is traced out by the reference state. This reference state is a vector that depends on the initial state vector of the system. In order to define this special open curve $\Gamma_{0}$, lets construct ``reference section" $|\chi(t)\rangle$ of the bundle covering  $\rho(t)=\Pi(\psi(t))$. It is a map $s:\mathcal{P}\xrightarrow{}\mathcal{L}$ such that the image of each point $\rho(t)\in\mathcal{P}$ lies in the fiber  $\Pi(\rho)$ over $\rho$, i.e., $\Pi o s = id_{p}$, for details see Ref. \cite{Isham}. The ``reference section" defined with respect to the initial point is a mapping of the state curve $\Gamma_{0}$ through the section s and is 
given by \cite{AKPati04,AKPati05} 
\begin{equation}
    |\chi(t)\rangle = \frac{\langle\psi(t)|\psi(0)\rangle}{|\langle\psi(t)|\psi(0)\rangle|}|\psi(t)\rangle.
\end{equation}
It has following properties: (i) $s\Pi(|\psi(0)\rangle)=|\chi(0)\rangle=|\psi(0)\rangle$, 
(ii) $\Pi(|\psi(0)\rangle)=\Pi(|\chi(0)\rangle)$, 
(iii) $\langle\chi(0)|\chi(t)\rangle$ is always real and positive, i.e., $|\chi(0)\rangle$ and $| \chi(t)\rangle$ remain in phase throughout the quantum evolution. 
Also, it insures that the length of the curve traced by $|\chi(t)\rangle$ is invariant under $U(1)$ gauge transformation. The ``reference-section" defined above plays an important 
role in the theory of geometric phases. Using this we can prove that the geometric phase acquired by a quantum system for an arbitrary non-cyclic evolution is given by the integral \
over the connection-form, i.e., $\Phi_G = i \int \langle \chi|d \chi \rangle$ \cite{AKPati04,AKPati05}.

Now, we will need two geometric structures to prove the reverse quantum speed limit. 
Consider two curves $\Gamma_0: [0, t] \rightarrow {\cal L} $ and ${\bar \Gamma}: [0, t] \rightarrow {\cal L} $ as traced out by the `reference 
 Section' $| \chi(t) \rangle$ and the horizontal curve $| {\bar {\psi}}(t) \rangle$, respectively. To define the later, let us consider the evolution of a quantum system as described 
 by the 
 Schr\"odinger equation
 $$  i\hbar\frac{d}{dt}|\psi(t)\rangle= H(t)|\psi(t)\rangle,$$
 where $H(t)$ is the driving Hamiltonian of the system. In general, when the system evolves in time from $t=0$ to $t=T$, it will acquire a dynamical phase
 $\Phi_D = - \frac{1}{\hbar} \int_0^T i \langle \psi(t)|H(t)| \psi(t) \rangle ~dt$ and a geometric phase $\Phi_G = \Phi - \Phi_D = i \int \langle \chi|d \chi \rangle $, where
 $\Phi = Arg \langle \psi(0)| \psi(T) \rangle $ is the total phase. However, 
 if the system undergoes a parallel transport then it will acquire only the geometric phase. The significance of the parallel transported vector is that locally it does not undergo 
 any rotation, but globally it picks up a phase which is geometric in nature. The parallel transported vector is given by 
 $| {\bar {\psi}}(t) \rangle = 
 \exp(i/\hbar \int_{0}^{t} \langle\psi(t')|H(t')| \psi(t')\rangle~ dt' ) |\psi(t)\rangle $ and this is also called as the horizontal vector. We can check that it 
 satisfies the parallel transport condition $\langle \psi(t)|\dot{\psi}(t)\rangle=0$, i.e., the tangent vector is orthogonal to the vector itself at any point in time.

Now, we need to define length of the curves for these two vectors. The inner product in ${\cal H}$ induces a metric in 
 ${\cal P}$ and the presence of metric allows the definition of the length of a differentiable curve in ${\cal L}$.\\

{\bf Definition (Length of reference section):}  Let $t \rightarrow | \chi(t) \rangle$ be a curve $\Gamma_0(t)$ during an arbitrary evolution of a quantum system. The total length of the differentiable curve $\Gamma_0$ from a point $| \chi(0) \rangle$ to a point $| \chi(t) \rangle$ is a real number defined as 
 \begin{align}
l{(\chi(t))}|_{0}^{T}& = \int_{0}^{T}{\langle\dot{\chi
}(t)|\dot{\chi}(t)\rangle}^\frac{1}{2}dt,\\
&=\int_{\Gamma_{0}}||d\chi||, \nonumber
\end{align}
where $|\dot{\chi}(t)\rangle$ is the velocity vector in $\mathcal{L}$ of the curve $\Gamma_{0}$ at time t along the path of evolution
(relative to the initial point $|\chi(0)\rangle$).

 {\bf Definition (Length of horizontal curve):}  Let $t \rightarrow | {\bar \psi(t)}  \rangle$ be a curve ${\bar \Gamma(t)} $ during an arbitrary evolution of a quantum system. The total length of the differentiable curve ${\bar \Gamma(t)} $ from a point $| {\bar \psi(0) }  \rangle$  to a point $| {\bar \psi(t)}  \rangle$ is a real number defined as 
 \begin{align}
l{({\bar \psi(t)})}|_{0}^{T}& = \int_{0}^{T}  {\langle \dot{{\bar \psi} }(t)| \dot{ {\bar \psi} }(t)\rangle}^\frac{1}{2}dt,\\
&=\int_{\bar \Gamma} ||d\bar\psi||, \nonumber
\end{align}
where $|\dot{{\bar \psi } }(t)\rangle$ is the velocity vector in $\mathcal{L}$ of the curve $\Gamma$ at time t along the path of evolution of the horizontal curve.\\

%................................................................
\begin{tikzpicture}

\draw[black,thick] (-3.8,-1) -- (2.5,-1,0) -- (3.8,1,0) -- (-2.5,1,0) -- cycle;

 \draw[black] (-2,1.5) -- (-2,4.4);
\draw[dotted,thick] (-2,0) -- (-2,1.5);

\draw [blue,dotted,thick] (2,0) arc[x radius=2, y radius=.5, start angle=0, end angle=180];

\draw[blue,thick] (-2,0) arc (-180:60:2 and 0.5);

\draw[rotate around={10:(-2,2)}][red,thick](-2,2) arc[x radius=2, y radius=.5, start angle=-180, end angle=0];

%\draw[rotate around={10:(-2,2)},violet,dotted,thick](-2,2) arc[x radius=2, y radius=.7, start angle=180, end angle=0];

x radius=2, y radius=.5, start angle=-180, end angle=50];

\draw[rotate around={10:(-2,2)}][red,thick]  (2,2) arc[x radius=2, y radius=.7, start angle=0, end angle=55];

%\draw[rotate around={-0.2:(-2,2.69)},red,dotted,thick]  (-2,2.69) arc (0:121:-2 and 0.5);

\draw[rotate around={20:(-2,2)}][green,thick](-2,2) arc[x radius=2, y radius=.5, start angle=-180, end angle=10];

\draw[rotate around={20:(-2,2)}][green,thick]  (2,2) arc[x radius=1.8, y radius=.7, start angle=0, end angle=49];

\node[black] at (-0.4,-1.25) {Projective Hilbert space  $\mathcal{P}$};

\node[black] at (-3.3,3) {Ray space $\mathcal{R}$};
%\node[black] at (-3.3,2.5) {(ray space)};

\node[black] at (1.95,3.5) {$\Gamma_0$};
%\node[black] at (2.4,3.6) {$|\chi(t)\rangle$};

\node[black] at (2.1,2.3) {$\bar \Gamma$};
%\node[black] at (2.4,1.6) { $| {\bar {\psi}}(t) \rangle$};

\node[black] at (-2.5,0) { $t=0$};
\node[black] at (1.3,0.7) { $t=T$};
\node[black] at (0,-0.75) { $\rho(t)$};
%\node[black] at (-0.4,0.75) { $\rho(t)_{Geodesic}$};
\node[thick,black] at (2.2,0) { $\hat{\Gamma}$};
\filldraw[blue] (1.0,0.425) circle (1pt) ;

\filldraw[blue] (-2,0) circle (1pt) ;
%\filldraw[red,thick] (-2,2.69) circle (1pt) ;
\filldraw[black,thick] (-2,1.98) circle (1pt) ;
\filldraw[red,thick] (1.0,3.12) circle (1pt) ;
\filldraw[green,thick] (1.0,3.65) circle (1pt) ;

\end{tikzpicture}

%................................................................
\begin{small}
Fig 2: In this schematic diagram the blue line represents the curve $\hat \Gamma$ in $\mathcal{P}$ which is projection of $|\psi(t)\rangle$ (as described in fig 1). The reference section curve $\Gamma_0$ and horizontal curve $\bar \Gamma$ are represented by green and red lines respectively. $\Gamma_0$ and  $\bar \Gamma$ traced by unit vectors $| { {\chi}}(t) \rangle$ and $| {\bar {\psi}}(t) \rangle$. These two curves do not depends on actual curve $\Gamma$ traced by $|\psi(t)\rangle$ and they are gauge invariant. 
\end{small} \\

%................................................................

The length of the reference curve and the length of the horizontal curves are two fundamental geometric structures associated with any quantum evolution. Some properties of these length of the curves are in the order: First, we note that the integrals in (3) and (4) exist in the interval $[0, T]$, since the integrand is continuous and the resulting integrals yield real numbers. These two lengths respect an important property of reparametrization invariance, i.e., all curves deduced from  $\Gamma_0$ and ${\bar \Gamma}$ by a change of parameter 
 $t$ to $t'$ with $\frac{dt}{dt'} > 0 $, the length of these curves remain unaltered. Furthermore, they are also gauge invariant, i.e., when 
 $| \psi(t) \rangle \rightarrow e^{i \alpha(t)} | \psi(t) \rangle$, then $l (\chi(t))$ and $ l(  {\bar \psi(t)}  ) $ remain the same. 
 This was proved in Ref. \cite{AKPati05}. Thus, they qualify to be called as geometric structures as these lengths are also independent of the particular Hamiltonian used to 
 evolve the quantum system. There may be infinite number of Hamiltonian which can give rise to same $l (\chi(t))$ and $ l(  {\bar \psi(t)}  ) $. 
 The length of the horizontal curve is actually (up to a factor 2) the total distance travelled by the quantum state as measured by the Fubini-Study metric.

To see this, consider the Bargmann angle, which measures the distance between two arbitrary pure states 
$|\psi_1\rangle$ and $|\psi_2\rangle$, is given by
\begin{equation}
 \frac{ 1}{2} S_o(|\psi_1\rangle,|\psi_2\rangle)= \cos^{-1}(|\langle \psi_1 |\psi_2 \rangle|).
\end{equation}
If two pure states $|\psi(t)\rangle $ and $|\psi(t+dt)\rangle $ are separated by an infinitesimal distance, then we have the infinitesimal Fubini-Study metric on $\mathcal{P}$, which is defined as \cite{Anandan02} 
 \begin{equation}
     dS^2 =  4 ( 1-| \langle \psi(t) | \psi(t+dt) \rangle|^2).
 \end{equation}
 Let $|\psi(t)\rangle$ is state of the system that evolves according to the Schr\"odinger equation
%\begin{equation}
%    i\hbar\frac{d}{dt}|\psi(t)\rangle= H(t)|\psi(t)\rangle,
%\end{equation}
%where $H(t)$ is the driving Hamiltonian. 
the, the distance between $|\psi(0)\rangle$ and  $|\psi(T)\rangle$ along the evolution curve is determined by integrating the Fubini-Study metric \cite{Anandan02} which is given by
\begin{equation}
    S  =  \frac{2}{\hbar}  \int_{0}^{T}  { \Delta H(t)} ~~dt,
\end{equation}
where $\Delta H(t)^2  =  \langle \psi(t)| H(t)^2|\psi(t)\rangle- \langle\psi(t)| H(t)|\psi(t)\rangle^2 $ is the energy fluctuation during the quantum evolution. 
Note that the length of the horizontal curve for the Schr{\"o}dinger evolution is given by $l{(  {\bar \psi(t)}  ) }|_{0}^{T} = \int_{0}^{T}  \frac{ \Delta H(t)}{\hbar} ~dt$,
i.e., $S=2l{(  {\bar \psi(t)}  ) }|_{0}^{T}$. The standard quantum speed limit (QSL) follows from the fact that the total distance travelled by the quantum system as measured by 
the Fubini-Study metric is always greater than or equal to the shortest distance connecting the initial and the final points, i.e., $S \ge S_0$. Similarly, here we will show 
how the geometry of quantum state space dictates that there is indeed a reverse quantum speed limit.

One fundamental result in geometry of quantum evolution is that the length of the reference section is greater than the length of the horizontal
 curve. In fact, we can prove that $d l(\chi(t)) ^2 \ge d l( {\bar \psi}(t)) ^2 $ and hence
the gauge invariant length $ l{(\chi(t))}|_{0}^{T}  $ is always greater than the length of the horizontal curve $l{(  {\bar \psi(t)}  ) }|_{0}^{T} $.

%................................................................
We can write Eq(2) as $|\chi(t)\rangle= \xi(t)|\psi(t)\rangle$, where $\xi(t)=\frac{\langle\psi(0 )|\psi(t)\rangle}{|\langle\psi(0 )|\psi(t)\rangle|}$. Then Eq(3) can be expressed as
%dl^{2}{(\chi(t))}&=[\frac{\xi^{*}(t)}{dt}\frac{\xi(t)}{dt}+2\frac{\xi^{*}(t)}{dt}\xi(t) {\langle{{ \psi} }(t)| \dot{ { \psi} }(t)\rangle}\\
%& + {\langle \dot{{ \psi} }(t)| \dot{ { \psi} }(t)\rangle}]dt^{2}.
\begin{align*}
\begin{split}
 ||d{\chi}||^{2} &=||{d\xi}||^{2}+2{d\xi^{*}}\xi {\langle{{ \psi} }| d{ { \psi} }\rangle} + ||{ d{ { \psi} }}||^{2}
\end{split}
\end{align*}
On using Eq(4) and the fallowing expressions
\begin{align*}
   ||{d\xi}||^{2} = [i{\langle {{ \chi} }| d{ { \chi} }\rangle} -i{\langle{{ \psi} }| d{ { \psi} }\rangle} ]^{2}, \\
    d\xi^{*}\xi =[{\langle{{ \psi} }| d{ { \psi} }\rangle} - {\langle {{ \chi} }| d{ { \chi} }\rangle}],
\end{align*}
we obtain
\begin{equation*}
   ||d{\chi}||^{2} -||d{\bar{\psi}}||^{2}=[i{\langle {{ \chi} }| d{ { \chi} }\rangle}]^{2}.
\end{equation*}
Since $i{\langle {{ \chi} }| d{ { \chi} }\rangle}$ is real, we have 
 %dl^{2}{(\chi(t))} \geq dl^{2}{(\bar \psi(t))}.
 
\begin{equation*}
\begin{split}
~~~~~~~~||d\chi||^{2} \geq ||d{\bar \psi}||^{2},  \\
 i.e., ~~~~~~l{(\chi(t))}|_0^T  &\geq l{(\bar \psi (t))}|_0^T.
 \end{split}
\end{equation*}

%................................................................

 The difference between the length and the distance plays a significant role which is essentially the connection-form that gives rise to the geometric phase for arbitrary quantum evolution \cite{AKPati04,AKPati05}.  Viewed differently, the existence of intrinsic curvature in quantum state space gives rise to the inequality, i.e., $ l{(\chi(t))}|_{0}^{T}   \ge l{(  {\bar \psi(t)}  ) }|_{0}^{T} $.

If the Hamiltonian is time-independent, then the above conditions provides a non-trivial  bound for the reverse 
speed limit which can be expressed as an inequality
\begin{equation}
   T  \le \frac{ \hbar  l{(\chi(t))}|_{0}^{T} }{\Delta H }.
\end{equation}
This is the fundamental reverse quantum speed limit (RQSL). If the speed of transportation of state vector is slow and if the total length of the 
``reference-section'' curve is more, then the system will evolve more slowly. 

For the time-dependent Hamiltonian, we can obtain the reverse speed limit for the quantum system as given by
\begin{equation}
  T  \le \frac{ \hbar  l{(\chi(t))}|_{0}^{T} }{ \overline{ \Delta H } },
\end{equation}
where ${\overline{ \Delta H }}$ is the time-average of the fluctuation over the time for which evolution occurs, i.e., ${\overline{ \Delta H }} = 
\frac{1}{T} \int_{0}^{T}  \Delta H(t) ~dt$. Thus, Eq(9) provides the reverse speed limit bound, i.e., upper bounds of speed limit for time as given by
\begin{equation}
   T_{RQSL} = \frac{ \hbar l{(\chi(t))}|_{0}^{T}}{\Delta H }.
\end{equation}

Since $l{(\chi(t))}|_{0}^{T} \geq l{(  {\bar \psi(t)}  ) }|_{0}^{T} \geq  \frac{1}{2}S_o(|\psi(0)\rangle,|\psi(T)\rangle) $, we can write the following inequality 
\begin{equation}
     T_{RQSL} \geq T\geq T_{QSL}.
\end{equation}
Thus, the geometric structures of the quantum evolution imposes fundamental bound on the evolution time as it is upper bounded by reverse speed limit time. 
Eq(12) suggests that in quantum  mechanics evolution  time  is  both upper  and lower bounded. It is worth stressing that the standard quantum speed limit as well as the reverse 
quantum speed limit obtained here, both owe their existence to the geometry of quantum state space. 
Eq(9) and (10) constitute the central result of our paper.

\subsection{For mixed initial state}
The reverse speed limit can be generalised for mixed initial states undergoing unitary time evolution. 
Any mixed state of a quantum system can be viewed as a reduced state of an enlarged pure entangled state. We can purify mixed state $\rho_{S}$ in $ \mathcal{H}_{S}$ of a system $(S)$ by attaching an ancillary system $(A)$. Hence state of the enlarged system $(S+A)$ described by pure state $|\Psi\rangle_{SA} \in \mathcal{H}_{S}\otimes \mathcal{H}_{A}$ and by tracing out $A$, we retrieve the mixed state $\rho_{S} \in \mathcal{B}(\mathcal{H}_{S})$ of $S$. The purified state is given by
\begin{equation}
    |\Psi\rangle_{SA}=\sum_{k}\sqrt{p_{k}}|k\rangle_{S}|k\rangle_{A},
\end{equation}
where $\{|k\rangle_{S}\}$ and $\{|k\rangle_{A}\}$ are basis of system Hilbert space $\mathcal{H}_{S}$ and ancillary system Hilbert space $\mathcal{H}_{A}$, respectively.

If $|\Psi(0)\rangle_{SA}$  is the initial state of joint system at time $t = 0$ and it is transformed to $|\Psi(t)\rangle_{SA} $ by a local unitary operator
$U_{SA}(t) = U_{S}(t) \otimes I{_A}$  \cite{D.Mondal54,ES55,D.Mondal54} then, we have
\begin{equation}
    |\Psi(t)\rangle_{SA}=\sum_{k}\sqrt{p_{k}}U_{S}(t)|k\rangle_{S}|k\rangle_{A},
\end{equation}
where $U_{S}(t)=e^{ \frac{-i t }{\hbar } H_{S} } $ and $H_{S} $ is the Hamiltonian of  the system. 
This unitary evolution is equivalent to  $\rho_S(0) \rightarrow \rho_S(t) = U_S(t) \rho_S(0) U_S(t)^{\dagger}$ as 
during the evolution at each time $t\in[0,T]$,
 $|\Psi(t)\rangle_{SA}$ satisfies the condition
\begin{equation}
    Tr_{A}[|\Psi(t)\rangle_{SA} \langle \Psi(t)| ] = \rho_{S}(t) = U_{S}(t)\rho(0)U_{S}(t)^{\dagger}, 
\end{equation}
where $\rho_{S}(0)$ is initial state of the system and $Tr_{A}[|\Psi(0)\rangle_{SA} \langle \Psi(0)| ] = \rho_{S}(0)$. 

Since, the purification is not unique, any state $  |\Psi(t)\rangle_{SA}=\sum_{k}\sqrt{p_{k}}U_{S}(t) | k \rangle_{S} V | k \rangle_{A}$,  
where $ V \in {\cal H}_A $  denotes a unitary operator in ${\cal H}_A $, is also a valid purification. Geometrically, this can be thought of as a right action of
 $V({\cal H}_{A})$  on $\mathcal{H}_{S}\otimes \mathcal{H}_{A}$ along the fibres
of the projection from  $\mathcal{H}_{S}\otimes \mathcal{H}_{A}$ to the space of density operators. This
projection is uniquely characterized by the equality
$ Tr ( \rho_{S} O_S )  =   \langle \Psi(t) |( O_S \otimes I_A) |\Psi(t) \rangle_{SA}$
satisfied by every operator $O$.  This can be regarded as a principal fibre bundle $\mathcal{H}_{S} \otimes \mathcal{H}_{A} $ 
over mixed states  with a structure group $V( {\cal H}_{A} )$  and a  well defined principal connection for mixed states \cite{L.dabrowski,A.Uhlmann}. 

In the purified Hilbert space $\mathcal{H}_{S} \otimes \mathcal{H}_{A} $, the entangled  ``reference section" with respect to initial state is defined as \cite{ES55,Du}
\begin{equation}
    |\chi(t)\rangle_{SA} = \frac{Tr(U_{SA}(t)\rho_{SA}(0))}{|Tr(U_{SA}(t)\rho_{SA}(0))|}|\Psi(t)\rangle_{SA},
\end{equation}
where $\rho_{SA}(0)=|\Psi(0)\rangle_{SA} \langle \Psi(0)|$ and $U_{SA}(t)=e^{ \frac{-i t }{\hbar } H_{S} } \otimes I{_A}$.

Now, if we carry over the notion of reference section and horizontal curve to the joint system. The length of reference section can be defined as 
\begin{eqnarray}
\label{eq2}
l{(\chi(t)_{SA})}|_{0}^{T} = \int_{0}^{T}  { \langle  \dot{\chi}(t)| \dot{\chi}(t) \rangle_{SA} }^\frac{1}{2}dt.
\end{eqnarray}
where $|\dot{\chi}(t)\rangle_{SA}$ is the velocity vector of the curve ${\Gamma}_{0}$ at time t along the path of evolution the path of evolution 
(relative to the initial point $|\chi(0)\rangle_{SA}$) in $\mathcal{H}_{S}\otimes \mathcal{H}_{A}$.

The length of the horizontal curve can be defined as 
\begin{eqnarray}
\label{eq}
l{(  {\bar \Psi(t)_{SA}}  ) }|_{0}^{T} = \int_{0}^{T}  {\langle \dot{{\bar \Psi} }(t)| \dot{ {\bar \Psi} }(t)\rangle_{SA} }^\frac{1}{2}dt,
\end{eqnarray}
where $|\dot{{\bar \Psi } }(t)\rangle_{SA}$ is the velocity vector of the curve $\bar{\Gamma}$ at time t along the path of evolution of the horizontal curve in $\mathcal{H}_{S}\otimes \mathcal{H}_{A}$.

During the evolution of the mixed state, we will have 
$ l{(\chi(t)_{SA})}|_{0}^{T}   \ge l{(  {\bar \Psi(t)}_{SA}  ) }|_{0}^{T} $. Now, realising the fact that 
$l{(  {\bar \Psi(t)}_{SA}  ) }|_{0}^{T} = Tr (\rho_S(t)  H_S^2) -  Tr (\rho_S(t) H_S )^2 = 
\Delta H_S^2$, we have the fundamental upper bound of quantum speed limit time  for mixed states which can be expressed as
\begin{equation}
   T \le \frac{ \hbar l{(\chi(t)}_{SA})|_{0}^{T} }{\Delta H_S }.
   \end{equation}
Indeed, we can check that the reverse quantum speed limit for mixed states given in Eq(19) reduces to the pure state case given in Eq(9), if the system is initially prepared in a pure state and undergoes a unitary time evolution.
    
%..........................................................................................................................

\section{ RQSL FOR THE TWO LEVEL QUANTUM SYSTEMS}
In the dynamics of two-level quantum system, we can easily examine the reverse speed limit bounds. Simplest models of two level systems are spin half particles in time dependent external field and the Jaynes-Cummings model. In the sequel, we will illustrate the reverse speed limit bounds for these two quantum systems.
 
%.................................................................................  

{\it Spin in time-dependent external field.--}Consider the atom (which has two energy levels) in time dependent external field whose Hamiltonian is given as
\begin{equation*}
    H = H_{atom} + H_{field}(t) = J_{1}\sigma_{z} + J_{2}(t)\sigma_{x},
\end{equation*}
where $J_{2}(t)$ is defined as $J_{2}(t) = 0 $ for $ t = 0$ and $J_{2}$ for $ t > 0$.
Here,  $|0\rangle$ and $|1\rangle$ are the excited and ground states of the atom with eigenvalues $\pm J_{1}$.
 If initial state of system is $|0\rangle$, then time evolution of the state of the system  at an arbitrary time is given by
\begin{equation*}
   |\psi(t)\rangle =[\cos(\frac{\Omega}{\hbar}t)-\frac{iJ_{1}}{\Omega} \sin(\frac{\Omega}{\hbar}t)]|0\rangle -\frac{iJ_{2}}{\Omega} \sin(\frac{\Omega}{\hbar}t)|1\rangle,
\end{equation*}
where $\Omega = \sqrt{{J_{1}}^2+{J_{2}}^2}$. The fluctuation in the energy of the  system during quantum evolution is given by
$\Delta H = J_{2}$.
In time interval $[0,T = \frac{\pi \hbar}{2 \Omega}]$, system reaches to target state  $\frac{J_{1}|0\rangle + J_{2}|1\rangle}{{{\sqrt{J_{1}^2 +  J_{2}^2}}}}$ (for simplicity, lets assume $J_{1}=J_{2}=\hbar$, it implies $T=\frac{\pi}{2\sqrt{2}}$). In order to obtain speed limit bounds,  we need to calculate $\frac{S_{0}}{2}$, $l{(\bar{\psi}(t))}|_{0}^{T}$ and $l{(\chi(t))}|_{0}^{T}$. We find that 
$ S_o(|\psi(0)\rangle,|\psi(T)\rangle) = \pi$, 
and $ l{(\bar{\psi}(t))}|_{0}^{T}= \frac{J_{2}}{\hbar}\frac{\pi}{2 \sqrt{2}} = 1.1107$. For the calculation of the length of the curve 
`$l{(\chi(t))}|_{0}^{T}$', first we need $|\chi(t)\rangle$. This is given by
\begin{equation}
\begin{split}
|\chi(t)\rangle=\frac{ \cos(\frac{\Omega}{\hbar}t)+\frac{iJ_{1}}{\Omega} \sin(\frac{\Omega}{\hbar}t)}{\sqrt{ \cos^2(\frac{\Omega}{\hbar}t)+(\frac{J_{1}}{\Omega})^2 \sin^2(\frac{\Omega}{\hbar}t)}} [[ \cos(\frac{\Omega}{\hbar}t) \\
-\frac{iJ_{1}}{\Omega} \sin(\frac{\Omega}{\hbar}t)]|0\rangle
 -\frac{iJ_{2}}{\Omega} \sin(\frac{\Omega}{\hbar}t)|1\rangle ] .\nonumber
 \end{split}
 \end{equation}
Using  Eq(3), we find the length of the reference curve as given by
\begin{align*}
l{(\chi(t))}|_{0}^{T}=
\frac{1}{\hbar}\int_{0}^{T} \sqrt{{\Omega}^2 +J_{1}^2 \frac{1-2b^2-2(1-b^2)\cos^2(at)}{(\cos^2(at)+b^2 \sin^2(at))^2}}dt, 
\end{align*}
where $a=\frac{\Omega}{\hbar}$ and $b=\frac{J_{1}}{\Omega}$. We can simplify the above expression further by substituting the values of  $J_{1}$ and $J_{2}$. This leads to

\begin{equation}
l{(\chi(t))}|_{0}^{T} = \int_{0}^{\frac{\pi}{2\sqrt{2}}} \sqrt{2- 4\frac{\cos^2(\sqrt{2}t))}{(\cos^2(\sqrt{2}t)+1)^2}}dt.  \nonumber
\end{equation}
The value of above integral is $ l{(\chi(t))}|_{0}^{T}= 1.2526 $ and 
$  l{(\chi(t))}|_{0}^{T}>l{(\bar{\psi}(t))}|_{0}^{T}>\frac{S_{0}}{2}$  indeed holds.
Using these values of $l{(\chi(t))}|_{0}^{T}$ and $\frac{S_{0}}{2}$,  the reverse speed limit and the standard speed limit are given by
\begin{align*}
T_{RQSL}=\frac{ \hbar  l{(\chi(t))}|_{0}^{T}}{\Delta H} = 1.2526, \\
T_{QSL}=\frac{ \hbar S_{0}(T)}{2\Delta H} = 0.7853.
\end{align*}
Thus, we have obtained desired upper and lower speed limit bounds on evolution time of given system, which completely agree with Eq(12).

 As of now we have obtained speed limit bounds for pure initial state. Let us consider the case, when initial state of the system is a mixed state, i.e., 
 $\rho_{S}=p|0\rangle \langle 0|+(1-p)|1\rangle \langle 1|$ with $0 \leq p \leq 1$.  After the purification, the state of system and ancillary is given by
 \begin{equation*}
    |\psi(0)\rangle_{SA} = \sqrt{p}|0\rangle_S | a_0 \rangle_A + \sqrt{1-p}|1\rangle_S |a_1  \rangle_A,
\end{equation*}
where $|a_0 \rangle$ and $|a_1\rangle$ are orthonormal basis of the ancillary system. Time evolution of joint system at a later time t is given by
\begin{equation*}
    |\Psi(t)\rangle_{SA} = U_{{SA}}(t) |\Psi(0)\rangle_{SA},
\end{equation*}
where  $U_{SA}= e^{\frac{ - it}{\hbar} (J_{1}\sigma_{z} + J_{2}(t)\sigma_{x}) } \otimes I_{A}$. The joint state at time t is given by
\begin{align*}
\begin{split}
    |\Psi(t)\rangle_{SA} =  \sqrt{p}[(\cos(\frac{\Omega t}{\hbar}) 
    -\frac{iJ_{1}}{\Omega}\sin(\frac{\Omega t}{\hbar}))|0\rangle \\
    - \frac{iJ_{2}}{\Omega}\sin(\frac{\Omega t}{\hbar})|1\rangle]| a_0 \rangle 
    + \sqrt{1-p}[-\frac{iJ_{2}}{\Omega}\sin(\frac{\Omega t}{\hbar})|0\rangle \\
    + (\cos(\frac{\Omega t}{\hbar})  +\frac{iJ_{1}}{\Omega}\sin(\frac{\Omega t}{\hbar}))|1\rangle  ]|a_1 \rangle,
\end{split}
\end{align*}
where $\Omega=\sqrt{{J_{1}}^2+{J_{2}}^2}$. The fluctuation of the Hamiltonian for the system  is given by 
$\Delta H_{S}  = \sqrt{J^2_{1}+4p(1-p)J^2_{2}} $.

Now, the reference section $|\chi(t)\rangle_{SA}$ can be expressed as 
\begin{equation*}
|\chi(t)\rangle_{SA} =\frac{\cos(\frac{\Omega}{\hbar}t)+\frac{iJ_{1}(2p-1)}{\Omega}\sin(\frac{\Omega}{\hbar}t)}{\sqrt{\cos^2(\frac{\Omega}{\hbar}t)+(\frac{J_{1}(2p-1)}{\Omega})^2 \sin^2(\frac{\Omega}{\hbar}t)}} |\Psi(t)\rangle_{SA}.
\end{equation*}
In time interval $[0,T=\frac{\pi \hbar}{2\Omega}]$, the system reaches to the target state  
$|\Psi(T)\rangle_{SA} $ 
(for simplicity, let us  assume that $J_{1}=J_{2}=\hbar$ and this implies $T=\frac{\pi}{2\sqrt{2}}$) with 
\begin{equation*}
\begin{split}
|\Psi(T)\rangle_{SA}=  \sqrt{p} |+ \rangle  |a_0\rangle 
    + \sqrt{1-p} |- \rangle  | a_1 \rangle,
\end{split}
\end{equation*}
where $\ket{\pm} = \frac{1}{\sqrt{2}}(|0\rangle   \pm  |1\rangle)$.
The state of the system at time $T$ is given by
\begin{equation*}
\begin{split}
    \rho_{S}(T)  =\frac{1}{2}(|0\rangle \langle 0| + |1\rangle \langle 1|) + (p -\frac{1}{2})(|0\rangle \langle 1|+|1\rangle \langle 0|).
    \end{split}
\end{equation*}
In order to obtain the reverse speed limit bounds, first we need to calculate the length of the curve.
For the purpose of illustration, we assume $p=\frac{1}{3}$ and the length is given by
\begin{equation*}
l{(\chi(t)_{SA}})|_{0}^{T} = \int_{0}^{\frac{\pi}{2\sqrt{2}}} \sqrt{2 - \frac{8 (1 + 17 \cos(2 \sqrt{2} t))}{(19 + 17 \cos(2 \sqrt{2} t))^2}}dt.
\end{equation*}
The value of above integral is 
$l{(\chi(t)_{SA}})|_{0}^{T} = 2.2458 $. Therefore, the reverse quantum speed limit bound is given by
$T_{RQSL}=\frac{ \hbar l{(\chi(t)_{SA}})|_{0}^{T} }{\Delta H_{S} } = 1.6341$. We also find that quantum speed limit and reverse speed limit respect the bound. Thus, the desired upper and lower speed limit bounds on evolution time for  mixed initial state completely agree with our new bound.

%.........................................................................

{\it The Jaynes-Cummings model.--}One may ask how tight is the reverse quantum speed limit? Is there any physical system for which RQSL saturates the bound? We will show that 
the Jaynes-Cummings (JC) model \cite{Jaynes-Cummings59} which describes the interaction of a two-level atom with a single quantized mode of an optical cavity’s electromagnetic field indeed saturates the reverse quantum speed limit. The Hamiltonian of JC model with rotating wave approximation can be expressed as
\begin{align*}
    H & = H_{atom}+H_{field}+H_{int}(t) \nonumber \\
& = \frac{\hbar\omega}{2} \sigma_{z} +\hbar\omega a^{\dagger}a + \lambda(t)(\sigma_{+}a + \sigma_{-}a^{\dagger}),
\end{align*}
where $\lambda(t)$ is defined as $  \lambda(t)  =    0  $  at  $ t = 0$ and  $ \lambda$ for  $ t > 0 $.
Here,  $|e\rangle$ and $|g\rangle$ are  the excited and ground states of the atom with eigenvalues $\frac{\hbar \omega}{2}$ and $-\frac{\hbar \omega}{2}$,  respectively. The cavity has `$n+1$' number of photons. If initial state of total system $|g\rangle |n+1\rangle$, then time evolution of the total system at arbitrary time t is given as
\begin{equation*}
|\psi(t)\rangle =\cos(\lambda t \sqrt{n+1} )|g\rangle|n+1\rangle
- i \sin(\lambda t  \sqrt{n+1} )|e\rangle|n\rangle,
\end{equation*}
where $|n\rangle$ and $|n+1\rangle$ are states of the field. The  energy fluctuation of the system during evolution is given by $\Delta H = \lambda \hbar \sqrt{n+1} $. Since initially atom in ground state $|g\rangle$, then in time interval $[0,T  = \frac{\pi}{2\lambda\sqrt{n+1}}]$, it evolves to target state $|e\rangle$. In order to obtain speed limit bounds for evolution, let us evaluate $\frac{S_{0}}{2}$, $l(\bar{\psi}(t))|_{0}^{T}$ and $l({\chi}(t))|_{0}^{T}$.
One can check that geodesic distance $S_{0}={\pi}$. The total distance as measured by the horizontal curve during the time evolution is given by $l(\bar{\psi}(t))|_{0}^{T} 
=\frac{\pi}{2}$.

In fact, the system undergoes parallel transport  during the quantum evolution, i.e., it satisfies the condition 
$\langle \psi(t) | {\dot \psi(t)} \rangle = 0$. 
In this case, we find that $|\chi(t)\rangle = |\psi(t)\rangle =  |\bar{\psi} \rangle $ and hence the length, distance and the geodesic distance all are equal during evolution the quantum system. 
In this scenario quantum speed limit bounds on evolution time saturates, i.e., it completely satisfy Eq(12) as well, i.e., $ T_{RQSL} =T = T_{QSL}$.

%..........................................................................................................................

\section{Discharging of Quantum Batteries}
Future technology aims to develop strategies to store energy which can be later consumed by
quantum devices. This motivates to design quantum batteries by using quantum mechanical features. The simplest model of quantum batteries consist of an array of $ N$ two-level quantum systems \cite{X.Zhang07,Y.Y.Zhang08,D.Ferraro10,F.C.Binder11,F.Campaioli12,F.Campaioli58}. However, their charging and discharging procedures differ because they use 
different external field to charge and discharge quantum batteries.
Once a quantum battery is successfully charged we need to decouple it from
the external field. In this connection, it has been shown that how fast a quantum battery can be charged is governed by the quantum speed limit for the evolution of the system. In this section, we will show that the reverse quantum speed limit provides a fundamental lower bound on the average discharging power alike quantum speed limit provides upper bound on average charging/discharging power.

\textit{Charging and Discharging Protocol}: In order to deposit work (charging) into array of $N$-atoms or extract work (discharging) from array of $N$ atom,  we apply a time dependent external field $H(t)$ for period $[0,T]$ such that it must be reversible cyclic operation \cite{F.C.Binder11}. The Hamiltonian during this process is given by 
\begin{align*}
H_{0}\xrightarrow{} H_{0} + H(t)\xrightarrow{} H_{0} ,
\end{align*}
where $H_{0}$ is Hamiltonian of quantum battery.
The norm of total Hamiltonian must be less than $E_{max}$ \cite{F.Campaioli12}, i.e.,
\begin{align*}
 ||H_{0} + H(t)|| \leq E_{max},
\end{align*}
where $E_{max}$ is maximum energy eigenvalue difference/gap of $H_{0}$. There are two ways to charge and discharge, first one is charging and discharging each atom individually known as parallel protocol and second one is charge and discharge all the atoms together known as the collective protocol \cite{F.C.Binder11,F.Campaioli12,F.Campaioli58}. Although, charging and discharging process are similar but we are only interested in discharging process. Because while charging we always want to charge it fast but that is not the case during the discharging process. We always wants to extract energy according to our necessity. Then, the natural question is how slow a quantum battery can discharge? Of course we have to remember that we cannot extract more than the stored energy. Here, we will show that the fundamental 
reverse speed limit can answer the question how long it takes to discharge a quantum battery. 

\textit{Average Discharging Power}: The average discharging power is defined as
\begin{equation}
\begin{split}
\bar{P}&=\frac{1}{T}\int_{0}^{T}\frac{dW(t)}{dt}dt,\\
&=\frac{W(T)}{T},
\end{split}    
\end{equation}
where $W(T)$ is the 
ergotropy and $T$ is time required to discharge the QB which is lower bounded by QSL time and upper bounded by RQSL time. 
The ergotropy  is the amount of energy deposited in quantum systems or extractable from quantum system \cite{A.E.Allahverdyan14} and defined as
\begin{equation}
    W(T) =\langle\psi (T)| H_{0}|\psi(T)\rangle -\langle\psi(0)|H_{0}|\psi(0)\rangle,
\end{equation}
where $H_{0}$ is Hamiltonian of QB, $|\psi (0)\rangle$ and $|\psi (T)\rangle$ are initial and final state of the QB, respectively.
This is the  maximum work that can be extracted unitarily from a given quantum state with respect to the Hamiltonian $H_0$.

The reverse speed limit provides a non-trivial lower bound to the average discharging power, i.e., we have 
\begin{equation}
\bar{P}  \ge \frac{W(T) \Delta H  }{ \hbar l (\chi(t) ) |_{0}^{T} },
\end{equation}
where $l (\chi(t) ) |_{0}^{T}$ is length of reference section and  $\Delta H$ is the energy fluctuation during the quantum evolution.

We will illustrate the average discharging power of quantum battery for two cases.

\subsection{Harmonic and square wave discharging}
Consider an $ N$-independent batteries consisting of $N$  two-level atoms which we can discharge though classical harmonic field \cite{Y.Y.Zhang08,Y.Y.Yan09}. The total Hamiltonian  of discharging process is described as
\begin{equation*}
    H = \frac{\varepsilon}{2}\sum_{i=1}^{N}\sigma_{i}^{z} +\frac{A}{2}\cos(\omega t)\sum_{i=1}^{N}\sigma_{i}^{x},
\end{equation*}
where the first term in the Hamiltonian denotes array of $N$ two level atoms and the second term denotes the classical harmonic field. Here, $|e\rangle$ and $|g\rangle$ are exited and ground states of single atom battery with eigenvalues $\frac{\varepsilon}{2}$ and$-\frac{\varepsilon}{2}$, respectively.

The effective Hamiltonian in the rotating wave approximation can be written as \cite{Y.Y.Zhang08,Y.Y.Yan09}
\begin{equation*}
    \bar{H} =  \frac{\bar{{\varepsilon}}}{2}\sum_{i=1}^{N}\sigma_{i}^{z} + \bar{A}\sum_{i=1}^{N}\sigma_{i}^{x},
\end{equation*}
where $\bar{A} = \frac{A}{2}(1-\frac{\zeta}{\sqrt{N}})$ and $\bar{\varepsilon} =\varepsilon J_{0} (\frac{A}{\omega\sqrt{N}}\zeta)-\omega$ (effective detuning)]. Here $\zeta\in[0,1]$ is an undetermined parameter and $J_{0} (\frac{A}{\omega\sqrt{N}}\zeta)$ denotes the Bessel function of order zero \cite{Y.Y.Zhang08,Y.Y.Yan09}.

The above Hamiltonian is similar to $N$ batteries coupled with square wave charger/discharger. In the discharging process of $N$ batteries, individual atoms  discharge independently (parallel discharging). If the  initial state of a single quantum battery is $|e\rangle$, then during the  discharging process the time evolution of wave function of the single atom battery system at arbitrary time $t$ is given by
\begin{equation*}
   |\psi(t)\rangle =[\cos(\frac{\Omega_{R}}{2\hbar}t)-\frac{i\bar{\varepsilon}}{\Omega_{R}}\sin(\frac{\Omega_{R}}{2\hbar}t)]|e\rangle -\frac{i2\bar{A}}{\Omega_{R}}\sin(\frac{\Omega_{R}}{2\hbar}t)|g\rangle,
\end{equation*}
where $\Omega_{R}=\sqrt{\bar{\varepsilon}^2+4\bar{A}^2}$.
The energy fluctuation of the system during the evolution is given by $\Delta H  =  \bar{A} $ and hence it evolves with a speed $\frac{2\bar{A}}{\hbar}$.

During the time interval $[0,\frac{\pi \hbar}{\Omega_{R}}]$ initial state  $|e\rangle$ evolves to target state $|\psi(T)\rangle$.  In order to calculate maximum and minimum discharging time of quantum battery, first we need to calculate $\frac{S_{0}}{2}$,      $l(\bar{\psi}(t))|_{0}^{T}$ and $l({\chi}(t))|_{0}^{T}$.
The geodesic distance $\frac{S_{0}}{2}$ is given by
\begin{align*}
\frac{1}{2}S_{0}(|\psi(0)\rangle,|\psi(T)\rangle)=\cos^{-1}(|\cos(\frac{\Omega_{R}}{2\hbar}T)-\frac{i\bar{\varepsilon}}{\Omega_{R}}\sin(\frac{\Omega_{R}}{2\hbar}T)|).
\end{align*}
The total length of the horizontal curve is given by $l(\bar{\psi}(t))|_{0}^{T}= \frac{\bar{A}}{\hbar}T$.

For the calculate of length $l(\chi(t))|_{0}^{T}$, first we need the reference section $|\chi(t)\rangle$ which
 can be expressed as
\begin{align*}
|\chi(t)\rangle = \frac{\cos(\frac{\Omega_{R}}{2\hbar}t)+\frac{i\bar{\varepsilon}}{\Omega_{R}}\sin(\frac{\Omega_{R}}{2\hbar}t)}{\sqrt{\cos^2(\frac{\Omega_{R}}{2\hbar}t)+(\frac{\bar{\varepsilon}}{\Omega_{R}})^2 \sin^2(\frac{\Omega_{R}}{2\hbar}t)}}[[cos(\frac{\Omega_{R}}{2\hbar}t) \\ 
-\frac{i\bar{\varepsilon}}{\Omega_{R}}\sin(\frac{\Omega_{R}}{2\hbar}t)]|e\rangle 
-\frac{i2\bar{A}}{\Omega_{R}}\sin(\frac{\Omega_{R}}{2\hbar}t)|g\rangle ].
\end{align*}
Now, the length of the reference section curve is given by
\begin{small}
\begin{equation*}
l(\chi(t))|_{0}^{T}= \frac{1}{\hbar}\int_{0}^{T} \sqrt{\frac{{\Omega_{R}}^2}{4}+\frac{\bar{\varepsilon}^2(1-2b^2-2(1-b^2)\cos^2(at))}{4(\cos^2(at)+b^2 \sin^2(at))^2}}dt,
\end{equation*}
\end{small}
where $a=\frac{\Omega_{R}}{2\hbar}$ and $b=\frac{\bar{\varepsilon}}{\Omega_{R}}$. 

In the  parallel discharging protocol, QSL and RQSL of discharging $N$ atoms battery is $N$ times of QSL and RQSL of single atom battery, respectively.
Thus, the reverse speed limit bounds on  discharging time of $N$ atoms is defined as

\begin{align*}
T_{RQSL}=N\frac{ \hbar l({\chi(t)})|_{0}^{T}}{\Delta H}   
\end{align*}

\begin{figure}
    \includegraphics[width=.495\textwidth]{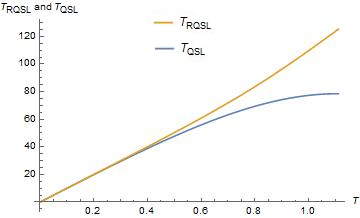}
{Fig 3: Here, we depict $T_{RQSL}$ and $T_{QSL}$ versus $T$ for $N=100$
atom battery for harmonic and square wave discharging. Here we have taken
$\Delta H$ in the unit of Joule-second; hence $T_{RQSL}$, $T_{QSL}$
and $T$ are in seconds.}
\end{figure}

The above Fig 3 shows that the upper and the lower bounds of discharging time of  hundred atoms ($N=100$) quantum battery. Here, we have assumed $\bar{A}=\hbar$ and $\bar{\varepsilon}=2\hbar$. In the plot, range of $T$ is $0$ to $\frac{\pi}{2\sqrt{2}}$, each value $T$ represents different final state $|\psi(T)\rangle$.
The total amount of work extracted from single atom battery during interval $[0,\frac{\pi \hbar}{\Omega_{R}}]$ is \cite{Y.Y.Zhang08}
\begin{equation*}
    W(T)=\frac{4\varepsilon \Bar{A}}{\Bar{\varepsilon}^{2}+4\Bar{A}^{2}}.
\end{equation*}
Therefore, the  upper and lower bounds on the average power of quantum battery are given by
\begin{equation*}
\frac{4\varepsilon \Bar{A}^{2}}{(\Bar{\varepsilon}^{2}+4\Bar{A}^{2})l({\chi}(t))|_{0}^{T}}  \leq \bar {P}\leq \frac{8\varepsilon \Bar{A}^{2}}{(\Bar{\varepsilon}^{2}+4\Bar{A}^{2})S_{0}}.
\end{equation*}
In time interval $[0,\frac{\pi \hbar}{2\bar{A}}]$ initial state  
$|e\rangle$ evolves to target state $|\psi(T)\rangle =  |g\rangle  $ (up to a phase), when we modulate $\omega$ such that  $ \bar{\varepsilon}=0 $ (tuned case). 
In this case $|\chi(t)\rangle =    |{\bar \psi(t)} \rangle$,
which implies that length, distance  and geodesic distance all are equal. This means that speed limit bounds saturates when $\bar{\varepsilon}=0$. Thus, the average power bound also saturates. Geometrically, for the tuned case, the system evolves along a shortest geodesic and obeys the parallel 
transport condition. However, for the detuned case, i.e.,  when $\bar{\varepsilon}\neq 0$ system may not evolve along geodesic and then quantum battery may take longer time to discharge. 
This can be harnessed in future to design efficient quantum batteries which will take more time to discharge. This shows that the geometry of quantum state space also dictates the 
discharging power of quantum battery.

\subsection{Cavity assisted discharging}
Consider the model of quantum battery \cite{X.Zhang07,D.Ferraro10,F.Campaioli12,Farina}, as array of $N$ two-level atoms inside the optical cavity. These atoms do not interact with each-other. The total Hamiltonian with rotating wave  approximation that describes the discharging process of the quantum battery is given by
\begin{equation*}
   H = \frac{\hbar\omega}{2}\sum_{j=1}^{N} \sigma_{z}^{j} +\hbar\omega a^{\dagger}a + \lambda(t)\sum_{j=1}^{N}(\sigma_{+}^{j}a + \sigma_{-}^{j}a^{\dagger}),
\end{equation*}
where the first term denotes the Hamiltonian of  $N$ atoms, the second term in Hamiltonian denotes single quantized mode of an optical cavity's electromagnetic field and the third term denotes the interaction between atoms and cavity \cite{Jaynes-Cummings59}. 
The cavity has $n$ number of photons, $\lambda(t) $ is a time-dependent coupling constant set to be $\lambda $  during  the  charging/discharging  period  $[0,T]$  and  $0$  otherwise.

 In the parallel discharging, we extract work form individual atoms independently using the external field. The time evolution of the single-atom battery system  is given by 
\begin{equation*}
{|\psi(t)\rangle =\cos(\lambda t \sqrt{n+1} )|e\rangle|n\rangle
- i \sin(\lambda t \sqrt{n+1}) |g\rangle|n+1\rangle} ,
\end{equation*}
where $|e\rangle$ and  $|g\rangle$ denotes exited and ground state of atom respectively. $|n\rangle$ and $|n+1\rangle$ are states of field. 
In time interval $[0,\frac{\pi}{2\lambda\sqrt{n+1}}]$ initial state of single atom $|e\rangle$ evolves to final state $|g\rangle$. In this model of quantum battery, we
find that the length of the reference curve, the length of the horizontal
 curve and the geodesic distance all are equal for discharging/charging of this quantum battery, i.e., $l({\chi}(t))|_{0}^{T}=l(\bar{\psi}(t))|_{0}^{T}=\frac{S_{0}}{2}$. 
 This  suggests  that reverse quantum speed limit bound saturates for this quantum battery model, thereby suggesting that such quantum battery takes fixed amount of time to 
 discharge. It also suggests that the lower and upper average power bounds also saturate during charging and discharging processes.\\

%.................................................................

\section{Conclusions}
In summary, we have proved a new reverse quantum speed limit for arbitrary unitary evolutions of pure as well as mixed states. 
The reverse speed limit arises due to geometry of the quantum state space, i.e., the total length of the reference curve is always greater than the length of the horizontal 
curve. The difference between these two lengths gives rise to the notion of curvature in the  state space.  Therefore, the reverse quantum speed limit owes its existence  due
to the intrinsic curvature on the projective Hilbert space of the quantum system. This is similar in spirit to the fact that the standard quantum speed limit also follows
from the geometric consideration, i.e., the total distance travelled by the quantum state as measured by the Fubini-Study metric is always greater or equal to the shortest
distance joining the initial and the final points on the projective Hilbert space of the quantum system.

Even though, the QSL and the RQSL bounds are fundamentally 
of geometrical nature, they differ in some important ways. First, to compute QSL one needs to know much less about
the path followed by the system than to compute RQSL. Second, evaluating QSL may be much easier than evaluating the actual time that the system takes to evolve from the initial to the final state, while evaluating the RQSL bound
may be difficult in physical situations. The QSL depends only on
the energy uncertainty and on the initial and final states, while the RQSL depends on the energy fluctuation and details of the path length followed
by the reference-state of the quantum system. Nevertheless, the RQSL will play an important role similar to the QSL.

We also find physical systems for which the upper bound for the reverse quantum speed limit
actually saturates. We have successfully presented  examples in support of our results. As an important application,
 we have shown how our result for reverse speed limit answers a pertinent question: how long it takes for a quantum battery to discharge? Our results
 shows that the geometry of quantum state space will play a key role in the future design of quantum battery. We have also shown that the
 reverse speed limit is tight by revealing  the cases when discharging  and charging time saturate in two different proposed models of quantum batteries.  In future, 
 we can generalize the reverse speed limit for open quantum systems and apply to the aging problem in quantum battery, i.e., what is the upper bound for the life time 
 of an open quantum battery? This will provide a route towards 
future extension and usability of our results in the context of battery stabilization.
 We believe that the fundamental reverse speed limit will have host of applications in quantum computing, quantum measurement, quantum control, quantum 
 metrology and variety of other areas.  

 \vskip 1cm
 
%\newpage
%................................................

\end{document}